\documentclass[10pt,journal,compsoc]{IEEEtran}

\usepackage{times}
\usepackage{epsfig}
\usepackage{graphicx}
\usepackage{amsmath}
\usepackage{amssymb}
\usepackage[accsupp]{axessibility}

\usepackage{tabularx}
\usepackage{amssymb}
\usepackage{booktabs}
\usepackage{color}
\usepackage{colortbl}
\usepackage{bbding}

\usepackage{multirow}
\usepackage[table,xcdraw]{xcolor}

\usepackage[pagebackref=true,breaklinks=true,colorlinks,bookmarks=false]{hyperref}
\usepackage{hyperref}

\definecolor{mygray}{gray}{.9}

\ifCLASSOPTIONcompsoc

  \usepackage[nocompress]{cite}
\else
  \usepackage{cite}
\fi

\ifCLASSINFOpdf

\else

\fi

\begin{document}

\title{Single-Image-Based Deep Learning for Segmentation of Early Esophageal Cancer Lesions}

\author{Haipeng~Li,~\IEEEmembership{Student Member,~IEEE,}
        ~Dingrui~Liu,
        ~Yu~Zeng,
        ~Shuaicheng~Liu, ~\IEEEmembership{Member,~IEEE,}
        ~Tao~Gan,
        ~Nini~Rao,
        ~Jinlin~Yang,
        and Bing~Zeng,~\IEEEmembership{Fellow,~IEEE}

\IEEEcompsocitemizethanks{

\IEEEcompsocthanksitem Manuscript submitted on June 09, 2023. This work was supported in part by the National Natural Science Foundation of China (NSFC) under Grant No. 61720106004.

\IEEEcompsocthanksitem Haipeng Li, Dingrui Liu, Shuaicheng Liu, and Bing Zeng are with the School of Information and Communication Engineering, Yu Zeng is with School of Glasgow College, Nini Rao is with School of Life Science and Technology, University of Electronic Science and Technology of China, Chengdu, Sichuan, 611731, China.

\IEEEcompsocthanksitem Tao Gan and Jinlin Yang are with West China Hospital, Sichuan University, Chengdu, China.

\IEEEcompsocthanksitem Haipeng~Li and Dingrui~Liu contributed equally.

\IEEEcompsocthanksitem Corresponding author: Bing Zeng (eezeng@uestc.edu.cn)
}
}

\IEEEtitleabstractindextext{%
\begin{abstract}
Accurate segmentation of lesions is crucial for diagnosis and treatment of early esophageal cancer (EEC). However, neither traditional nor deep learning-based methods up to today can meet the clinical requirements, with the mean Dice score - the most important metric in medical image analysis - hardly exceeding 0.75. In this paper, we present a novel deep learning approach for segmenting EEC lesions. Our approach stands out for its uniqueness, as it relies solely on a single image coming from one patient, forming the so-called "You-Only-Have-One" (YOHO) framework. On one hand, this "one-image-one-network" learning ensures complete patient privacy as it does not use any images from other patients as the training data. On the other hand, it avoids nearly all generalization-related problems since each trained network is applied only to the input image itself. In particular, we can push the training to "over-fitting" as much as possible to increase the segmentation accuracy. Our technical details include an interaction with clinical physicians to utilize their expertise, a geometry-based rendering of a single lesion image to generate the training set (the \emph{biggest} novelty), and an edge-enhanced UNet. We have evaluated YOHO over an EEC data-set created by ourselves and achieved a mean Dice score of 0.888, which represents a significant advance toward clinical applications. 
\end{abstract}

}

\maketitle

\begin{figure*}[t]
\begin{center}
  \includegraphics[width=1\linewidth]{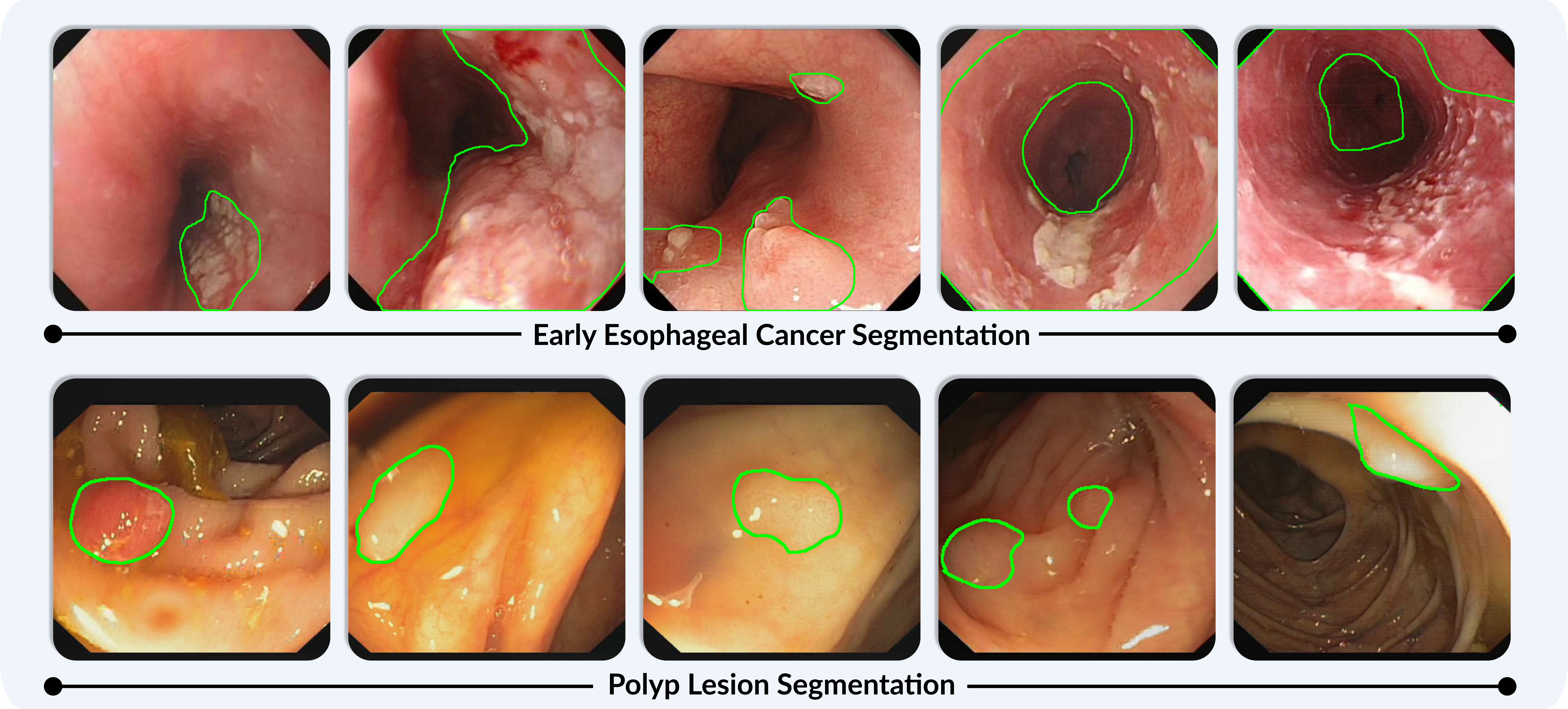}
\end{center}
  \caption{Five images in the first row demonstrate the complexity of EEC lesions in these individual cases so that accurate lesion segmentation would be highly challenging as compared to polyp lesions shown in the second row.}
\label{examples}
\end{figure*}

\IEEEdisplaynontitleabstractindextext

\IEEEpeerreviewmaketitle

\section{Introduction}
\label{sec:introduction}

Esophageal cancer is the sixth most common type of cancer worldwide and the eighth leading cause of cancer-related deaths, resulting in approximately 600,000 fatalities annually~\cite{fang2022identification,stein2005early}. However, early diagnosis and treatment can significantly reduce the mortality. In clinical practice, doctors often rely on endoscopy images to identify early esophageal cancer (EEC) lesions~\cite{pimentel2015endoscopic, oyama2005endoscopic}. The first row of Fig.~\ref{examples} shows some examples of EEC lesions where the green-colored contours for the ground-truth of lesion regions are hand-sketched by experienced endoscopists. Generally, EEC lesions vary greatly in individual cases: from left to right, 5 images in Fig.~\ref{examples} representing an easy case, a large lesion, multiple lesions, and circular lesions surrounded by two rings, respectively.

For the purpose of comparison, we show in the second row of Fig.~\ref{examples} some endoscopy images, chosen from the open data-set CVC-612~\cite{bernal2015wm}, that contain polyp lesions. It is clear that EEC lesions have much more complicated shapes, and therefore accurate segmentation of EEC lesions would be a lot more difficult.

Over the past two decades, computer-aided methods have been developed to segment various lesions in endoscopy images~\cite{sezgin2004survey,senthilkumaran2009image,achanta2012slic, chan2001active,li2010distance}. In particular, with the recent development of deep learning-based methods~\cite{ronneberger2015u, zhou2018unet++, fan2020pranet, cao2021swin, dong2021polyp, kirillov2023segment}, segmentation precision has gradually improved. For instance, several most advanced deep neural networks (DNNs) for polyp segmentation have achieved the mean Dice score higher than 0.90~\cite{fan2020pranet,cao2021swin,dong2021polyp}. However, these DNNs become much less accurate for EEC lesion segmentation, with mean Dice scores typically below 0.75, which is hardly acceptable in clinical applications. The reasons are twofold. First, as demonstrated above, EEC lesions are much more complex than polyps, making segmentation considerably more difficult. Second, medical data for EEC is very limited due to privacy protection and the required data annotation by experienced physicians, making the trained network prone to "over-fitting" and thus significantly decreasing its performance on new lesion images. 

In this work, we focus on the EEC lesion segmentation, with the goal of solving both the privacy-related issue and the up-bounded precision problem. In contrast to using a training set (composed of images obtained from multiple patients) in nearly all today's deep learning methods, we limit our algorithm to see images from a single patient only. In the extreme case, there is only one image available. This setting aligns with the clinical practice where endoscopists usually observe a single best image, taking into account factors like view-angle, lighting, and sharpness, to make the diagnosis.

Specifically, we propose the so-called "You-Only-Have-One" (YOHO) framework for deep learning, in which the learning is based solely on a single lesion image. Because of this "one-image-one-network" nature, the lesion image of a patient is used to serve himself/herself, thus providing a complete protection of the patient's pvivacy. In YOHO, a DNN is trained for each input lesion image individually and then applied to the same image. As there are no generalization-related issues, the training can be made "over-fitting" as much as possible. Consequently, YOHO has the potential to significantly improve the segmentation precision. 

Given a single endoscopy image $\boldsymbol{x}$ that contains pathological changes such as EEC (or polyp) lesions, YOHO consists of three steps: generate a diverse and qualitative training set (which is unique to $\boldsymbol{x}$), train a DNN based on this set, and apply the trained network to $\boldsymbol{x}$ to accomplish the lesion segmentation. Clearly, the first step, i.e., how to perform rendering from the single input image, is the most important one. To this goal, clinical endoscopists are first involved to help locate lesions in a polygon or several polygons and then take several circle-shaped samples over the lesion area. In this way, doctors are kept in the loop so as to fully utilize their expertise. Second, the sampled-circles are cut to form several geometric shapes (e.g., new circles and triangles) to be pasted back to $\boldsymbol{x}$ in randomly-selected positions. This cutting-pasting is repeated many times so as to create the training set.

Note that each data in such training set consists of several lesion instances of cut-and-pasted circles and triangles. The ground-truth information for these circles and triangles, including their masks and edges, is known. To train our model, we have selected UNet~\cite{ronneberger2015u} and put an additional emphasis on edges. Details of the training procedure will be presented in Section~\ref{sec:method}.

\section{Related Works}
\label{sec:related_work}

While there is a large number of works on medical image detection/segmentation, our discussions presented in this section focus on lesion detection/segmentation in endoscopy images, particularly for polyp and early esophageal cancer (EEC). We select polyp because there are open and widely-used datasets~\cite{bernal2015wm,jha2020kvasir} so that our method can be compared with the most recent works. We select EEC because it has a higher clinical value and for this task we have created a dataset by ourselves.

\subsection{Polyp Detection/Segmentation}
\label{sec:endoscopy}

Traditional approaches are more concentrated on detecting each polyp into a rectangular box~\cite{mamonov2014automated, tajbakhsh2015automated, zhang2016automatic} by extracting hand-crafted features, such as color, edge, texture, geometry, appearance, or a combination of these characteristics. Due to the limitation on traversing as many features as possible, they are prone to mis-detection of lesions. This is also one reason why they are not often used for the pixel-wise polyp segmentation. Similar scenarios happened also in several early attempts of applying deep learning-based methods~\cite{yu2016integrating, zhang2018polyp}, i.e., they are aiming at locating the lesion with a rectangular box (or circle). Recently, there appear a number of deep learning-based methods for the pixel-wise segmentation of polyps~\cite{brandao2017fully,akbari2018polyp}. In particular, the popular UNet~\cite{ronneberger2015u} and several variations such as UNet++~\cite{zhou2018unet++} and ResUNet++~\cite{jha2019resunet++} have been applied successfully to accomplish this task, demonstrating a better performance as compared to the traditional level-set method. 

Most recently, PraNet~\cite{fan2020pranet} employs attention modules for the polyp segmentation task; Swin-UNet~\cite{cao2021swin} introduces the swin transformer to model the long-range and global contextual information; and Polyp-PVT~\cite{dong2021polyp} adopts a transformer-based encoder to learn more powerful and robust representations. These networks have further improved the segmentation performance and represent the state-of-the-arts (SOTAs).

\subsection{EEC Segmentation}
\label{sec:ECC_segmentation}

Compared to polyp detection, detection of early esophageal cancer (EEC) is by far more important. In practice, as the EEC lesion usually has a complicated shape, it becomes more difficult to segment as compared to polyp segmentation. Very recently, there appear several works that propose the use of deep learning-based methods for this task~\cite{fang2022identification, le2020application, cai2019using, ebigbo2019computer, guo2020real}. Meanwhile, the deep learning-based methods mentioned above for polyp segmentation can also be applied to EEC segmentation. However, all these methods can hardly provide accurate-enough results, e.g., the mean Dice score remains at a low level.

In addition, it is important to point out that the accessible EEC database is currently not existing. To solve this problem, we have created our own EEC dataset. Details will be presented in Section~\ref{sec:dataset}.

\begin{figure*}[t]
\begin{center}
  \includegraphics[width=1\linewidth]{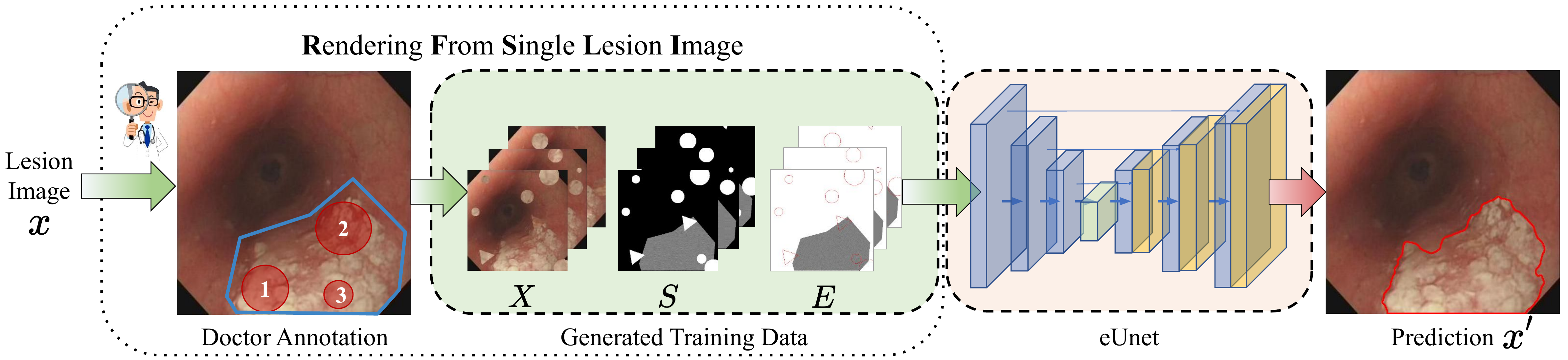}
\end{center}
  \caption{Pipeline of our proposed YOHO framework. Here, doctor's involvement at the beginning plays a critical role in avoiding segmentation errors (missed region and false-alarmed region) that might happen in the conventional deep learning methods, see Fig.~\ref{fig:CVC_qualitative_results},  Fig.~\ref{fig:EEC_qualitative_results}, and the supplementary material for some examples.}
\label{fig:overall_ppl}
\end{figure*}

\subsection{Weakly Supervised Interactive Segmentation}
\label{sec:zero_shot}

To overcome the reliance on dense pixel-level annotation, several weakly supervised interactive methods have been proposed for medical image segmentation. For instance, DeepCut~\cite{rajchl2016deepcut} uses bounding boxes as weak annotations for the training set and adopts an iterative strategy similar to Grabcut~\cite{rother2004grabcut}; DeepIGeoS~\cite{wang2018deepigeos} uses geodesic distance transforms of scribbles as additional channels for interactive segmentation; BIFSeg~\cite{wang2018interactive} incorporates CNNs into a bounding box and scribble based segmentation pipeline to deal with previously unseen object classes;  UGIR~\cite{wang2020uncertainty} refines the segmentation results based on the probabilistic uncertain region of the CNN's output. In addition, Zhang \emph{et. al.}~\cite{zhang2021interactive} proposed to enhance the segmentation by learning the bi-directional sequential patches guided by the given central point of the object.

Nevertheless, none of the above methods deals with endoscopy images. Furthermore, they all need datasets that are formed by collecting images from multiple patients and the involved interactions are relatively complex (e.g., it is required in multiple rounds during the training). In contrast, our approach works on one image obtained from the outpatient himself/herself and needs a simple interaction from physicians only at the beginning.

\section{Our Method}
\label{sec:method}

First, we would like to state that the study protocol has been approved by the Ethics Office of the Department of Clinical Research Management, West China Hospital, Sichuan University, and all participants have signed a written informed consent to take part in the study.

Given a single endoscopy image $\boldsymbol{x}$ that contains diagnosed lesions, our YOHO framework is illustrated in Fig.~\ref{fig:overall_ppl}, consisting of three steps as follows: 

\begin{itemize}

\item generate three datasets from $\boldsymbol{x}$: the lesion image set 
$\boldsymbol{X}=\left\{\boldsymbol{x}_1, \ldots, \boldsymbol{x}_K\right\}$, the segmentation mask set 
$\boldsymbol{S}=\left\{\boldsymbol{s}_1, \ldots, \boldsymbol{s}_K\right\}$, and the edge map set 
$\boldsymbol{E}=\left\{\boldsymbol{e}_1, \ldots, \boldsymbol{e}_K\right\}$, i.e., 
\begin{equation}
    \{ \boldsymbol{X}, \boldsymbol{S}, \boldsymbol{E} \} =\mathcal{R}\left(\boldsymbol{x}, 
\Pi_1\right),
\end{equation} where $\Pi_1$ represents the hyper-parameters involved in the generation procedure;

\item train a DNN based on three generated datasets: 
\begin{equation}
\{\hat{\boldsymbol{s}},\hat{\boldsymbol{e}},\hat{\boldsymbol{e}}^{\prime}\} =\underset{\Theta}{\arg \min } \mathcal{L}\left(\boldsymbol{X},\boldsymbol{S}, \boldsymbol{E},\Pi_2\right), 
\end{equation} where $\Theta$ stands for the trained network and $\Pi_2$ denotes the hyper-parameters involved in the training procedure. In addition, $\hat{\boldsymbol{s}}$ represents the segmentation output, $\hat{\boldsymbol{e}}$ and $\hat{\boldsymbol{e}}^{\prime}$ denote two edge outputs;

\item apply the trained DNN to $\boldsymbol{x}$ to accomplish the segmentation.

\end{itemize}

\begin{figure*}[t]
\begin{center}
  \includegraphics[width=1\linewidth]{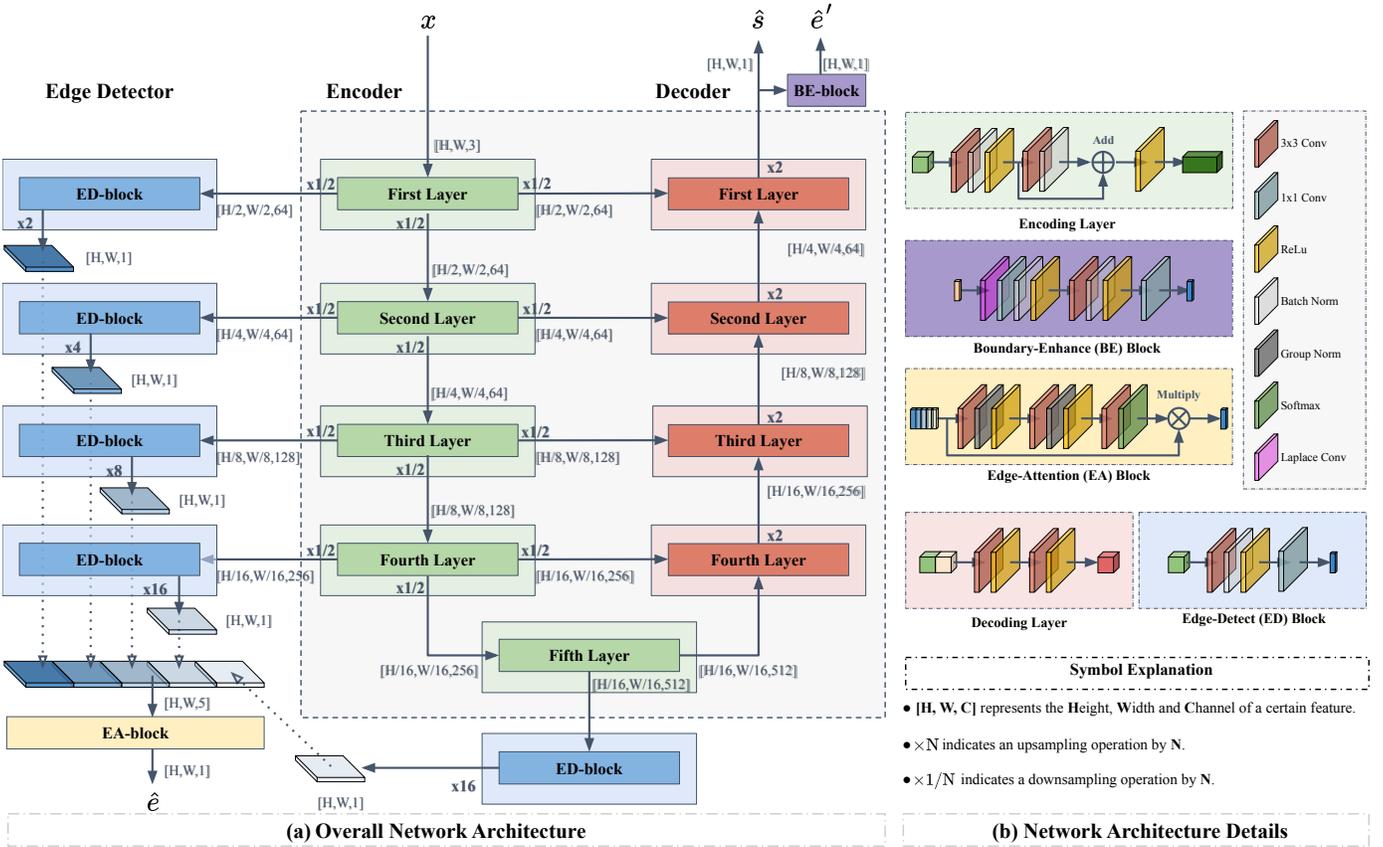}
\end{center}
\caption{Overall architecture of our eUNet: it is based on the popular UNet with a depth of 5, while the edge-detection (ED) module is added to extract edge information from encoding features. The right part shows some details of the key building blocks.}
\label{fig:eUnet_pipeline}
\end{figure*}

\subsection{Rendering from a single lesion image.}\label{sec:RFSL} Clinical endoscopists are involved to sketch the lesion area of the input image $\boldsymbol{x}$ by a polygon or several polygons, forming the region of interest (ROI). They are further required to take some samples over the lesion area. These samples are all circle-shaped with different diameters, forming a sample set: $\boldsymbol{C} = \left\{\boldsymbol{c}_1, \ldots, \boldsymbol{c}_N\right\}$ where $N$ is small.

All sampled circles are cut into new circles of different diameters and equilateral triangles of different orientations. These circles and triangles form another set: $\boldsymbol{G} = \left\{\boldsymbol{g}_1, \ldots, \boldsymbol{g}_M\right\}$, $M>>N$, which represents all lesion seeds. Then, we select several seeds from $\boldsymbol{G}$ randomly and paste them back to $\boldsymbol{x}$ at random positions but in the non-overlapping fashion. After repeating this process many times, we generate the training set: $\boldsymbol{X}=\{\boldsymbol{x}_1, \ldots, \boldsymbol{x}_K\}$ where $K>>M$. In the meantime, the segmentation mask set $\boldsymbol{S}$ and the edge map set $\boldsymbol{E}$ are generated automatically, which act as the ground-truth.

\vskip 0.15cm
\noindent\textbf{Reverse ROI.} It is difficult to sketch the ROI when the lesion occupies a majority area, which is common in the EEC dataset. To address this issue, we propose the reverse ROI. In contrast to ROI that outlines lesion areas, the reverse ROI is constructed by sketching polygons over the healthy region. In this way, the output of our network is the segmentation of the healthy region and the lesion area can be obtained by inverting the network’s output.

\subsection{Edge-enhanced UNet.} The overall architecture of our edge-enhanced UNet (eUNet) is shown in Fig.~\ref{fig:eUnet_pipeline}(a). The main part of eUNet (in the dashed box) is the conventional 5-layer UNet~\cite{ronneberger2015u}. We choose UNet as the backbone because of its proven effectiveness of producing good segmentation results (particularly for medical images)~\cite{ronneberger2015u,zhou2018unet++,jha2019resunet++}. Here, ResNet34 is used as the encoder and ReLU as the activation function. The output obtained after all decoding layers represents the segmentation result $\hat{\boldsymbol{s}}$. Furthermore, a boundary-enhance (BE) block is proposed to extract an edge map $\hat{\boldsymbol{e}}^{\prime}$ from the segmentation output $\hat{\boldsymbol{s}}$.

The other part of eUNet functions as an edge detector of 5 layers. The input at each layer comes from the feature map of the corresponding encoder. Each layer produces an edge map of the same resolution via an edge-detection (ED) block (an up-sampling is needed here). Then, five edge maps are concatenated and fed into an attention block to generate another edge map, denoted as $\hat{\boldsymbol{e}}$. 

Notice that our proposed eUNet produces two edge maps. It is easy to understand that estimating an accurate edge map helps the estimation of the segmentation mask and vice versa. By enforcing the consistency between two edge maps (one of them representing the edge of the segmentation mask), we expect that the final segmentation would be improved because its edge has been enhanced.

\subsection{Loss function.} We jointly learn the segmentation result $\hat{\boldsymbol{s}}$ and two edge maps $\hat{\boldsymbol{e}}$ and $\hat{\boldsymbol{e}}^{\prime}$ through the end-to-end network. The loss function consists of 3 parts as follows.

\vskip 0.15cm
\noindent\textbf{Segmentation Loss}. To learn the segmentation output, a combined loss of binary cross entropy (BCE) and Dice loss~\cite{milletari2016v} is adopted:

\begin{equation}
\small
\mathcal{L}_{\text {seg}}(\hat{\boldsymbol{s}}, \boldsymbol{s})=\mu_1 \mathcal{L}_{B C E}(\hat{\boldsymbol{s}}, \boldsymbol{s})+\mu_2 \mathcal{L}_{\text {Dice }}(\hat{\boldsymbol{s}}, \boldsymbol{s}),
\end{equation}

\noindent where $\hat{\boldsymbol{s}}$ is the segmentation output, $\boldsymbol{s}$ is the ground-truth segmentation mask, and $\mu_1$ and $\mu_2$ are weights.

\vskip 0.15cm
\noindent\textbf{Edge Loss}. As the proportion of boundary information in an edge map is very small, we use a weighted cross entropy (WCE) loss~\cite{liu2017richer} to handle the high imbalance between boundary and non-boundary pixels:

\begin{equation}
\small
\mathcal{L}_{\text {edge }}(\hat{\boldsymbol{e}}, \boldsymbol{e})= \text{WCE}(\hat{\boldsymbol{e}}, \boldsymbol{e}),
\end{equation}

\noindent where $\hat{\boldsymbol{e}}$ and $\boldsymbol{e}$ represent the output of the edge detector and the ground-truth edge map, respectively.

\vskip 0.15cm
\noindent\textbf{Consistency Loss}. To further enhance the edge information, an edge-consistency loss is proposed as follows: 

\begin{equation}
\small
\mathcal{L}_{\text {consist }}(\hat{\boldsymbol{e}}^{\prime}, \hat{\boldsymbol{e}})=\text {WCE}(\hat{\boldsymbol{e}}^{\prime}, \mathcal{o}(\hat{\boldsymbol{e}})),
\end{equation}

\noindent where $\hat{\boldsymbol{e}}^{\prime}$ is the output produced by the BE-block and $\mathcal{o}$(.) represents the binarization operation.

\vskip 0.15cm
Finally, the total loss function is constructed by a weighted sum of the three losses defined above:

\begin{equation}
\small
\mathcal{L}_{\text {total }}=\lambda_1 \mathcal{L}_{\text {seg }}(\hat{s}, s)+\lambda_2 \mathcal{L}_{\text {edge }}(\hat{e}, e)+\lambda_3 \mathcal{L}_{\text {consist }}(\hat{e}^{\prime}, \hat{e}),
\end{equation}

\noindent where $\lambda_1$, $\lambda_2$, and $\lambda_3$ are the weighting coefficients. 

Details about the involved weights and the parameters used in our training will be presented in Section~\ref{sec:implementation_details}.

\begin{table*}[t]
    \small
    \centering
    \resizebox{\linewidth}{!}{
\begin{tabular}{cl|cccccccc} \toprule 
\textbf{Training-set required} & \textbf{Model name}  & \textbf{mDice} & \textbf{mIoU}  & \textbf{$F_{\beta}^{w}$}  & \textbf{\quad $S_{\alpha}$}  & \textbf{$E_{\phi}^{\max }$}  & \textbf{MAE}  & \textbf{Recall}  &  \textbf{Precision} \\ \midrule 
                           
    $\checkmark$ & Unet~\cite{ronneberger2015u}   & 0.686                     & 0.546    & 0.631  & 0.641  & 0.740   & 0.182 & 0.736 & 0.723  \\
    
    $\checkmark$ & Unet++~\cite{zhou2018unet++}      & 0.690                      & 0.561    & 0.607  & 0.614  & 0.681   & 0.211   & n/a   & n/a  \\

    $\checkmark$ & PraNet~\cite{fan2020pranet}      & 0.683                     & 0.559    & 0.628  & 0.677  & 0.718   & 0.179 & 0.746 & 0.744  \\
    
    $\checkmark$ & Swin-Unet~\cite{cao2021swin}   & 0.715                      & 0.587    & 0.665  & 0.672  & 0.756   & 0.168   & 0.711   & 0.750  \\ 
    
    $\checkmark$ & Polyp-PVT~\cite{dong2021polyp}   & \underline{0.736}                      & \underline{0.611}    & \underline{0.681}  & \underline{0.684}  & \underline{0.759}   & \underline{0.164}  & \underline{0.810}  & \underline{0.752}  \\
    
     & \textbf{Ours} & \textbf{0.888}  & \textbf{0.804}    & \textbf{0.870}  & \textbf{0.839}  & \textbf{0.929}   & \textbf{0.059}  & \textbf{0.884}  & \textbf{0.903} \\  \bottomrule
\end{tabular} }  
\vspace{0.3cm}\caption{Eight quantitative metrics are computed on EEC-2022 for various methods.}
    \label{table:eec_quantity}
\end{table*}

\subsection{Discussions}

\begin{figure*}[h]
\begin{center}

  \includegraphics[width=1\linewidth]{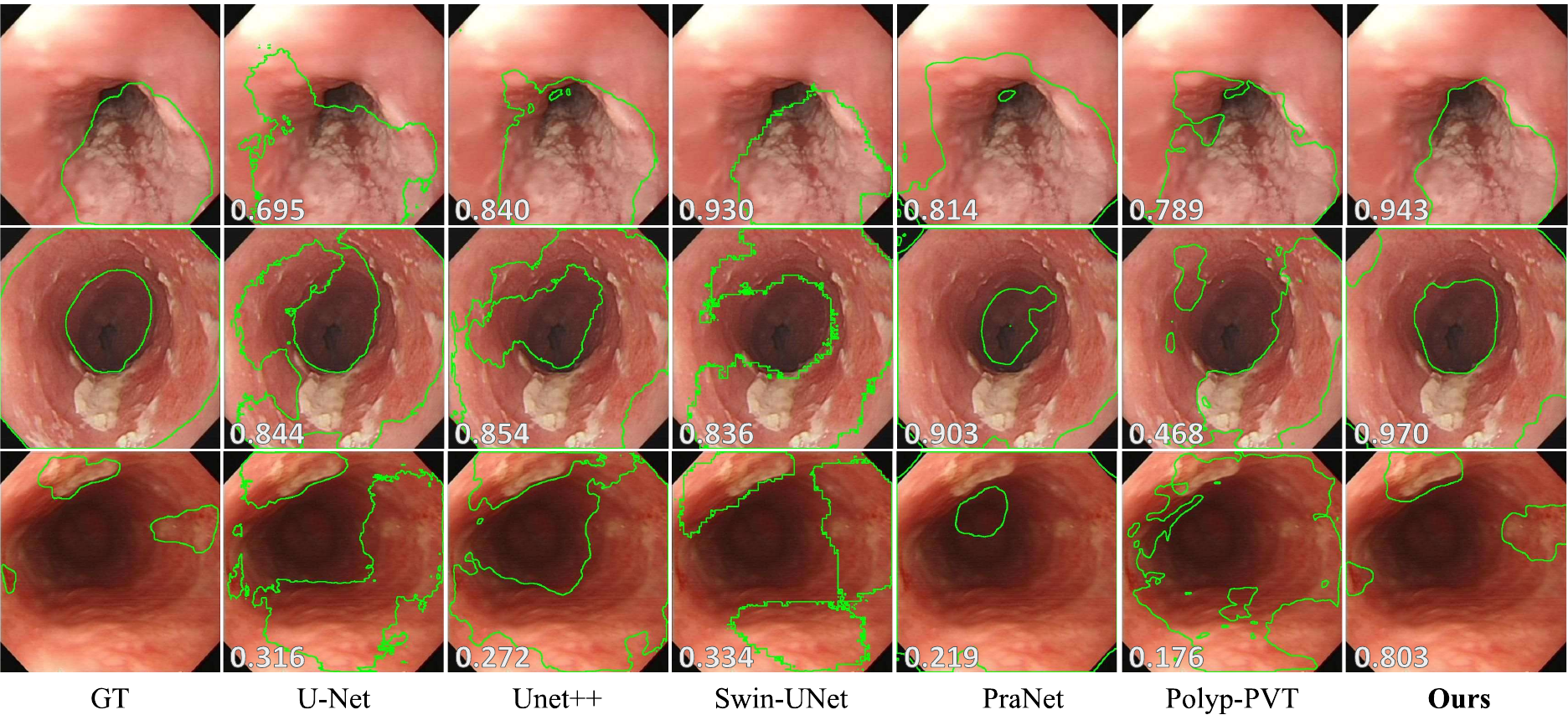}
\end{center}
  \caption{Some segmentation results of different methods on EEC-2022 (the Dice score of each result is shown at the bottom-left corner). More results will be presented in the supplementary material.}
\label{fig:EEC_qualitative_results}
\end{figure*}

\noindent To our best knowledge, the mechanism underpinning our YOHO framework is original. Firstly, it differs from the "one-shot" or "few-shot" learning method that requires one or a few samples for each class\cite{fei2006one,ravi2016optimization,snell2017prototypical}. Secondly, it is also different from the "zero-shot" learning that relies on the transfer of domain-specific knowledge\cite{palatucci2009zero,lampert2009learning}. Thirdly, although it seems similar to the pre-training in the popular "self-supervised" framework \cite{doersch2017multi,sun2021autoflow} (e.g., both start with unlabeled data and create pseudo but useful labels), the key difference is that the network trained in our framework is directly applied to the input image. In contrast, a fine-tuning that is based on some labeled data is typically required to adapt each specific task in the "self-supervised" framework.

Technically, we believe that the most significant innovation of our YOHO framework is its ability to render data from a single lesion image. Although we use only circles and triangles, we expect that once a DNN learns to detect these simple shapes, it will also be capable of fitting complex contours. This is because short pieces of an arbitrary contour can be approximated by a line or an arc, allowing the whole contour to be approximated by short lines and arcs in a piece-wise manner. We chose equilateral triangles because other types of triangles have sharper angles, which would make the training more difficult to converge.

\begin{figure*}[t]
\begin{center}
  \includegraphics[width=1\linewidth]{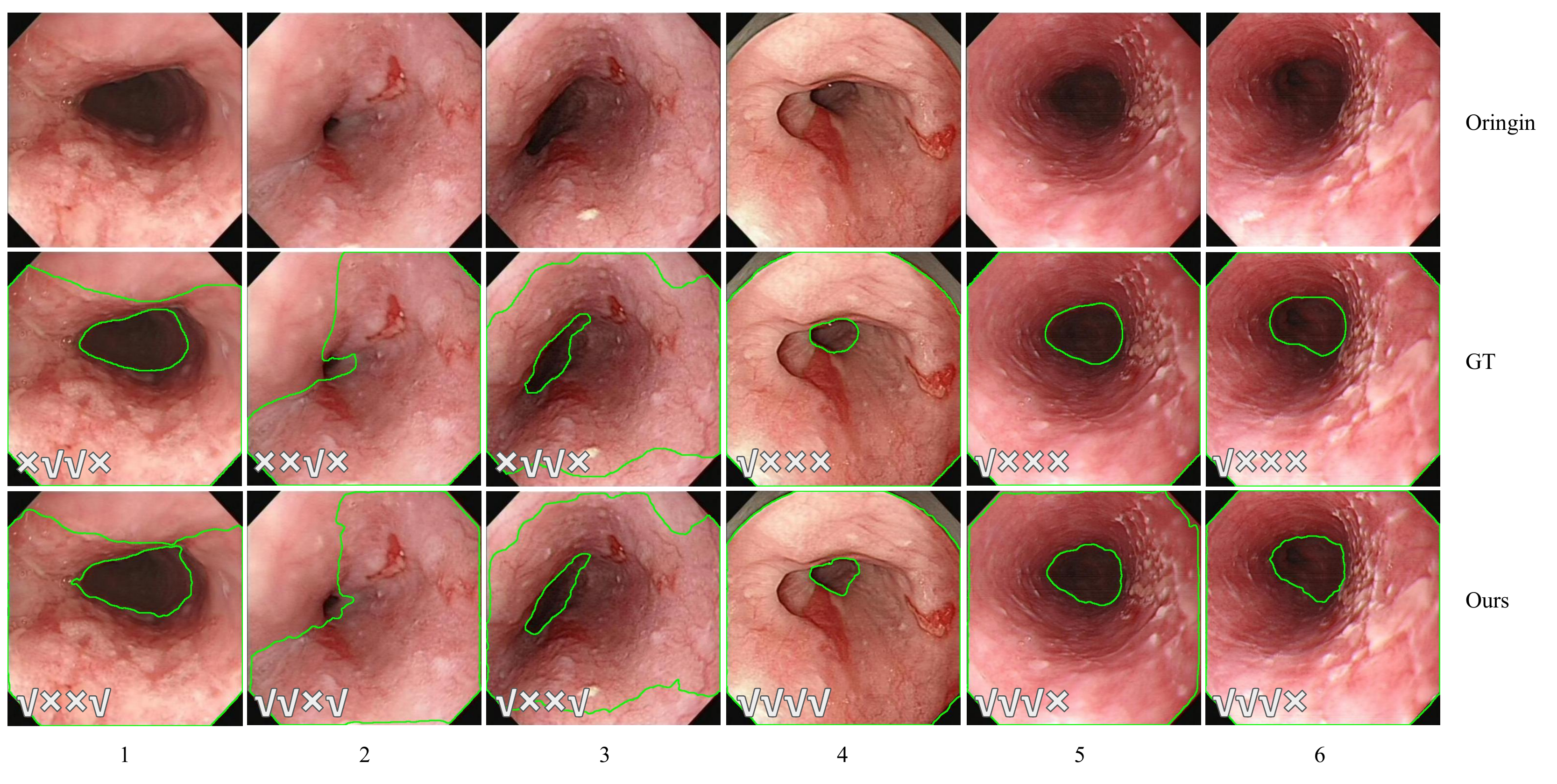}
\end{center}
  \caption{Some blind evaluations (1st - 6th case) by endoscopists. More results will be presented in the supplementary material.}
\label{fig:blind_eval_1}
\end{figure*}

\section{Experiments}
We first describe the datasets and implementation details. Then, we conduct comprehensive experiments to verify the effectiveness of our proposed framework, including quantitative comparisons, qualitative comparisons, user studies, generalization study, comparison with SAM, and ablation studies.

\subsection{Datasets}
\label{sec:dataset}

Three datasets are used in our experiments. The first one is the EEC dataset that is created by ourselves, and two of them are popular polyp datasets with open access.

\vskip 0.15cm
\noindent \textbf{Early Esophageal Cancer 2022 (EEC-2022)} is a dataset we created in this work. To the best of our knowledge, there is no dedicated dataset for EEC yet. Our EEC-2022 dataset is annotated by our students under very close supervision from experienced endoscopists. It consists of 1236 images of resolution $480 \times 480$, and is divided into the training set of 1092 samples and the test set of 138 samples.

\vskip 0.15cm
\noindent \textbf{CVC-ClinicDB}\cite{bernal2015wm} (also referred as CVC-612) is a standard polyp segmentation dataset extracted from colonoscopy videos. It consists of 612 samples of resolution $384 \times 288$ along with the ground-truth masks corresponding to the polyp region. The dataset is divided into the training set of 550 samples and the test set of 62 samples, exactly the same as what is done in PraNet~\cite{fan2020pranet}.

\vskip 0.15cm
\noindent \textbf{Kvasir-SEG}\cite{jha2020kvasir} is a larger polyp dataset annotated by a medical doctor and verified by an experienced gastroenterologist. It provides 1000 polyp images of resolution ranging from $332 \times 487$ to $1920 \times 1072$ along with the corresponding segmentation masks. Again, the same partitioning as used in PraNet is applied here to obtain the training set of 900 samples and the test set of 100 samples.




\begin{table*}[]
    \small
    \centering
    
    \resizebox{\linewidth}{!}{
\begin{tabular}{cc|cccccccc}
\hline  
{ \textbf{Data-set}} &{ \textbf{Training set required}} & { \textbf{Model name}}  & { \textbf{mDice}} & { \textbf{mIoU}}  & { \textbf{$F_{\beta}^{w}$}}  & { \textbf{\quad $S_{\alpha}$}}  & { \textbf{$E_{\phi}^{\max }$}}  & { \textbf{MAE}} \\ \hline

                                            \hline
                                            
                                            \multirow{5}*{\rotatebox{90}{\small \textbf{CVC-612}  \quad}} 

                                           &$\checkmark$ & Unet~\cite{ronneberger2015u}        & 0.823  & 0.755    & 0.811  & 0.889  & 0.954   & 0.019  \\
                                            
                                           &$\checkmark$ & Unet++~\cite{zhou2018unet++}      & 0.794                      & 0.729    & 0.785  & 0.873  & 0.931   & 0.022  \\
                                            
                                            
                                           &$\checkmark$ & PraNet~\cite{fan2020pranet}      & 0.899                     & 0.849    & 0.896  & 0.936  & 0.979   & 0.009  \\
                                            
                                           &$\checkmark$ & Swin-Unet~\cite{cao2021swin}   & 0.837                      & 0.763    & 0.830  & 0.887  & 0.932   & 0.015  \\ 
                                            
                                           &$\checkmark$ & Polyp-PVT~\cite{dong2021polyp}      & \underline{0.937}                     & \underline{0.889}    & \underline{0.936}  & \underline{0.949}  & \underline{0.985}   & \textbf{0.006}  \\
                                            
                                            
                                            && \textbf{Ours} & \textbf{0.940}  & \textbf{0.890}    & \textbf{0.949}  & \textbf{0.950}  & \textbf{0.988}   & \underline{0.008}  \\
                                            
                                            \hline
                                            \hline
                                            
                                            \multirow{5}*{\rotatebox{90}{\small \textbf{Kvasir} \quad}} 
                                            
                                           &$\checkmark$ & Unet~\cite{ronneberger2015u}        & 0.818  & 0.746    & 0.794  & 0.858  & 0.893   & 0.055  \\
                                            
                                           &$\checkmark$ & Unet++~\cite{zhou2018unet++}      & 0.821 & 0.743    & 0.808  & 0.862  & 0.910   & 0.048  \\ 
                                            
                                           &$\checkmark$ & PraNet~\cite{fan2020pranet}      & 0.898                     & 0.840    & 0.885  & 0.915  & 0.948   & 0.030  \\
                                            
                                           &$\checkmark$ & Swin-Unet~\cite{cao2021swin}   & 0.864                      & 0.791    & 0.858  & 0.882  & 0.927   & 0.035  \\ 
                                            
                                           &$\checkmark$ & Polyp-PVT~\cite{dong2021polyp}      & \underline{0.917}                     & \textbf{0.864}    & \underline{0.911}  & \underline{0.925}  & \underline{0.962}   & \underline{0.023}  \\
                                            
                                            
                                            && \textbf{Ours} & \textbf{0.924}  & \underline{0.862}    & \textbf{0.923}  & \textbf{0.926}  & \textbf{0.973}   & \textbf{0.019}  \\
                                            
                                            \hline
                                            
\end{tabular} }  
\vspace{0.3cm}\caption{Six quantitative metrics are computed on CVC-612 and Kvasir for various methods.}
    \label{table:polyp_quantity}
\end{table*}

\begin{figure*}[h]
\begin{center}
  \includegraphics[width=1\linewidth]{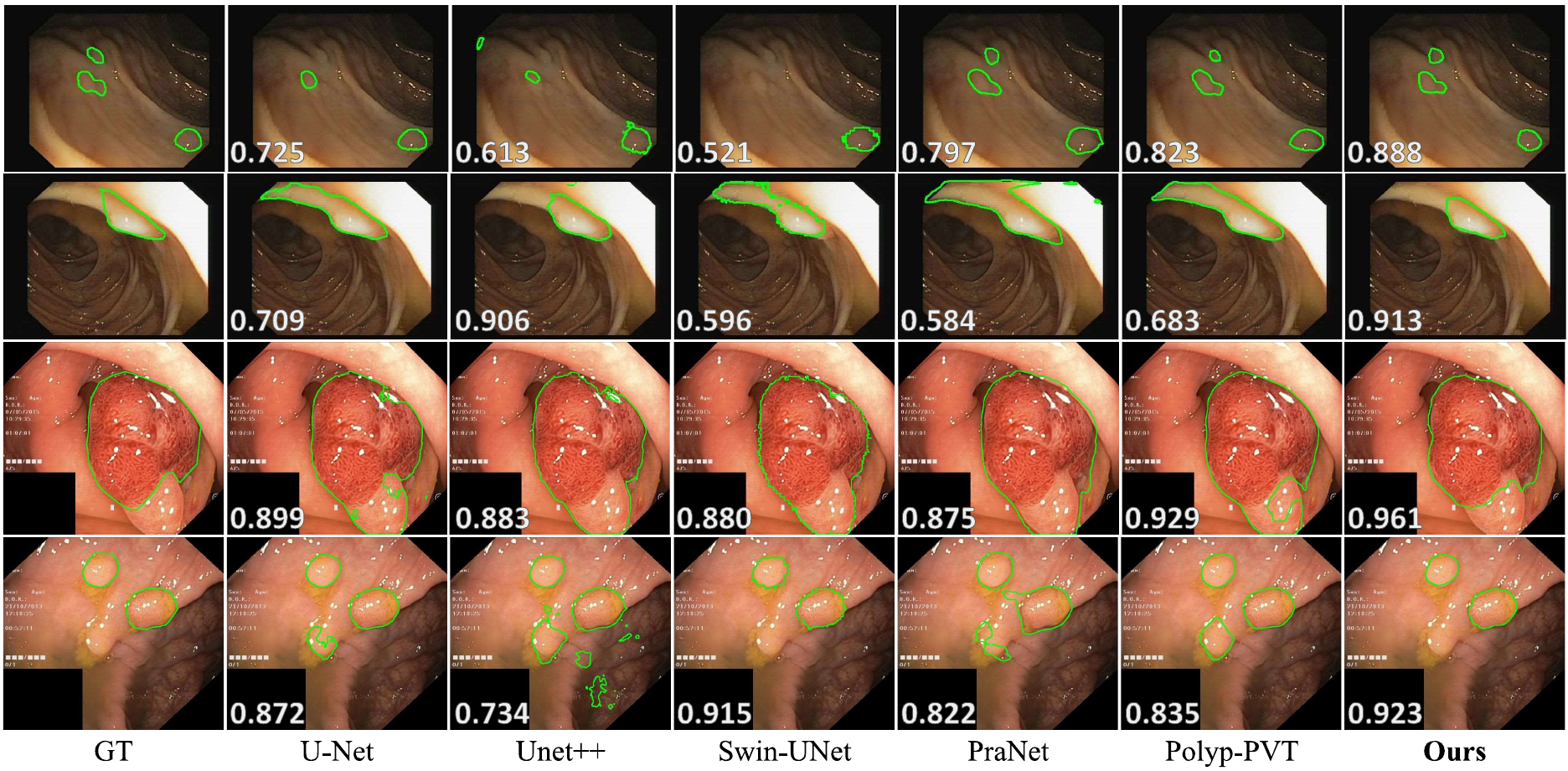}
\end{center}
  \caption{Some segmentation results of different methods on CVC-612 (1st and 2nd rows) and Kvasir (3rd and 4th rows).}
\label{fig:CVC_qualitative_results}
\end{figure*}

\subsection{Implementation Details}
\label{sec:implementation_details}


All deep learning networks (such as UNet~\cite{ronneberger2015u}, UNet++~\cite{zhou2018unet++}, PraNet~\cite{fan2020pranet}, Swin-UNet~\cite{cao2021swin}, and Polyp-PVT~\cite{dong2021polyp}) that are compared in our experiments are trained based on the partitioned training sets in CVC-ClinicDB, Kvasir, and EEC-2022. Then, each trained network is applied to the corresponding test set. 

Notice that to test an image $\boldsymbol{x}$ in the test set, $\boldsymbol{x}$ is fed as input into each trained network (UNet, UNet++, PraNet, Swin-UNet, or Polyp-PVT) so that the network "sees" $\boldsymbol{x}$. In YOHO, to process $\boldsymbol{x}$, our algorithm only sees $\boldsymbol{x}$ also. Therefore, the comparison between YOHO and other deep learning networks (i.e., UNet, UNet++, PraNet, Swin-UNet, and Polyp-PVT) is fair. The difference is that YOHO needs to do more steps: generate the unique training set from $\boldsymbol{x}$, train an eUNet, and apply the trained eUNet to $\boldsymbol{x}$.

To implement YOHO on the input image $\boldsymbol{x}$, our students first sketch a polygon or several polygons to cover the lesion area and then take 2-10 samples. They accomplish this task under close supervisions from experienced endoscopists. Note that each sketched polygon is quite rough, as shown in Fig.~\ref{fig:overall_ppl}. It is our expectation that an even better performance would be obtained if sketching and sampling are conducted directly by experienced endoscopists.

Then, YOHO starts to work: it first generates 1600 lesion images through the cutting and pasting as described earlier in Section.~\ref{sec:RFSL} and down-sizes each image to the resolution $256 \times 256$ to form the training set $\boldsymbol{X}$; eUNet is then trained over the training set; and finally the converged eUNet is applied to the input image $\boldsymbol{x}$ to accomplish the segmentation.

To train eUNet, we first freeze the ResNet34 module and train the full model for 1,000 steps with batch size 32 by an Adam optimizer~\cite{kingma2014adam} where the learning rate is set at $l_{r}=1.0 \times 10^{-3}$ and reduced by $10\%$ every 50 steps. Then, we unfreeze the ResNet34 module to refine the full model for 1,000 steps where the learning rate is set to $l_{r}=3.0 \times 10^{-4}$ and decays by $10\%$ every 50 steps. Other parameters involved in the training are described in the supplementary material. The total processing time is about 4.5 minutes on Intel i$9$ $12900$K CPU with $10,240$M memory. The implementation is in PyTorch and 2 NVIDIA RTX 2080 Ti are used.

\begin{figure*}[t]
\begin{center}
  \includegraphics[width=1\linewidth]{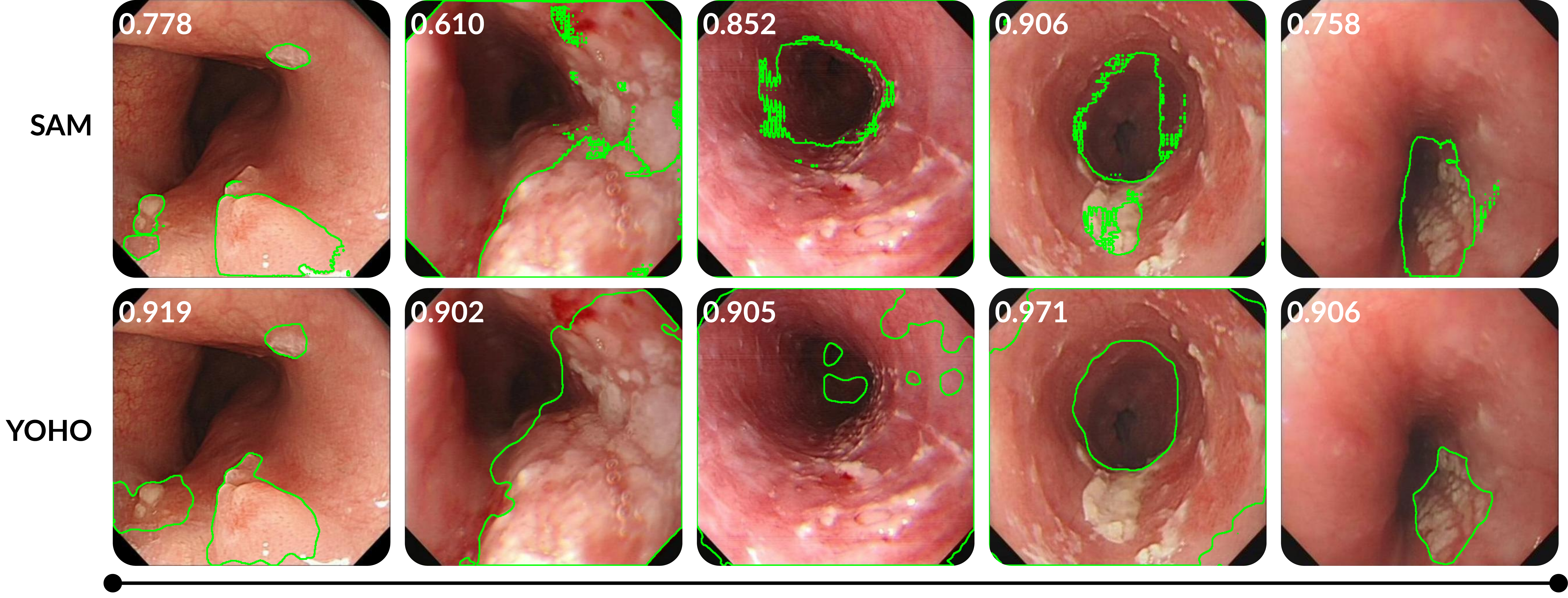}
\end{center}
  \caption{Comparisons between YOHO and SAM on several EEC lesions. The Dice score is shown on the left-top corner.}
\label{sam_teaser}
\end{figure*}
\subsection{Quantitative comparisons.} Table~\ref{table:eec_quantity} presents the quantitative comparisons between YOHO and the chosen benchmark models based on 8 metrics. The mean Dice score is most widely used in medical image analysis, while the mean IoU, recall, and precision are important metrics for image segmentation. In addition, we use the weighted Dice $F_{\beta}^{w}$ to correct "equally important defects". The MAE metric evaluates accuracy at the pixel level, while the enhanced alignment metric $E_{\phi}^{\max}$\cite{fan2018enhanced} assesses pixel-level and global-level similarity. Lastly, we use $S_{\alpha}$\cite{fan2017structure} to evaluate the similarity between predictions and the ground truths.

Table~\ref{table:eec_quantity} clearly shows that YOHO's performance gain is very significant in every metric, e.g., the mean Dice score increases by more than 10 percentage points.

\subsection{Qualitative comparisons.} Figure~\ref{fig:EEC_qualitative_results} presents some EEC lesion segmentation results. It can be seen that the ground-truth contours differ significantly from each other and thus may also be different from those encountered in the training set based on which those benchmark models are trained. Consequently, these benchmark models often fail in locating the lesion area accurately and produce highly non-smooth contours with jagged edges. In contrast, YOHO segments lesions quite smoothly and accurately.

\subsection{Blind evaluation by doctors.} We conducted a blind evaluation on 30 samples from EEC-2022, involving 4 experienced endoscopists. During the blind test, we presented each participant with a row of images, including the original lesion image in the middle, and our segmentation output and the ground-truth at random positions on the left and right. Each participant was asked to independently choose among the following options: (1) the left-side result is better, (2) the right-side result is better, (3) two are equally good, and (4) both are bad. They were not informed that one of these two results is the ground-truth, to ensure an unbiased evaluation.

Table~\ref{table:user_study} presents the choices made by 4 participants, demonstrating that our results are highly comparable to the ground-truth. Notably, Fig.~\ref{fig:blind_eval_1} illustrates some qualitative comparisons. Here, the first row shows the original lesion images, the second row displays the ground-truth labels, and the last row presents our segmentation results. Specifically, we use a check $\checkmark$ and a cross $\times$ to represent the voting results. For instance, "$\checkmark \checkmark \times \times$" shown in one image indicates that the first two doctors prefer this result, but the last two doctors vote against it. Our results appear to be slightly preferred by the doctors. Additionally, while each doctor's subjective judgment differs, they all tend to support the effectiveness of the results, with only 3 out of 120 instances being deemed ineffective.

\subsection{Generalization to polyp segmentation.} For polyp segmentation, we select two popular data-sets: CVC-612~\cite{bernal2015wm} and Kvasir~\cite{jha2020kvasir}. We follow the same partitioning of each data-set into the training set and the test set as used in PraNet~\cite{fan2020pranet}, i.e., 548/64 images in the training/test set in CVC-612 and 900/100 images in the training/test set in Kvasir, respectively. The quantitative results shown in Table~\ref{table:polyp_quantity} confirm that YOHO outperforms the chosen benchmark networks in nearly all metrics. In addition, the qualitative results shown in Fig.~\ref{fig:CVC_qualitative_results} demonstrate that the benchmark models produce some serious errors (missed or false-alarmed regions), while YOHO consistently performs well on all examples. This is largely due to the sketching of initial polygons by experienced physicians, which helps YOHO to avoid the missing or false-alarming problem.

\begin{table}[]
    \small
    \centering
    \resizebox{1\linewidth}{!}{
\begin{tabular}{c|cccccc}
\hline  
\multicolumn{1}{l}{} & \textbf{eUNet}  & Ground-Truth & Equally Good & Both Bad\\ \hline 

    Doctor 1 & 14 & 5 & 11 &0 \\
    Doctor 2 & 9 & 21 & 0 &0 \\
    Doctor 3 & 10 & 20 & 0 &0 \\
    Doctor 4 & 22 & 5 & 0 &3 \\
    Total & \textbf{55} & 51 & 11 &3 \\
    \hline
\end{tabular}    
}
    \vspace{0.3cm}\caption{Blind evaluation results of 4 experienced endoscopists where 30 samples from EEC-2022 are used.}
    \label{table:user_study}
\end{table}

\begin{table}[]
    \small
    \centering
    \resizebox{1\linewidth}{!}{
\begin{tabular}{ccc|cc}
\hline  
\multicolumn{1}{l}{} & Consistency Loss & Edge Detector & mean Dice  & mean IoU   \\ \hline 

    exp 1 &  &  & $0.906$ & $0.835$ \\
    exp 2 &  & $\checkmark$ & \underline{$0.909$} & \underline{$0.839$}  \\
    \textbf{Ours}& $\checkmark$ & $\checkmark$ & \textbf{0.940} & \textbf{0.890}\\
    
    \hline
\end{tabular}    
}
    \vspace{0.3cm}\caption{Comparative results without using the edge loss and the edge consistency loss.}
    \label{table:ablation_study_3}
\end{table}

%

\begin{table}[]
    \small
    \centering
    \resizebox{1\linewidth}{!}{
\begin{tabular}{c|cccccc}
\hline  
\multicolumn{1}{l}{} & mean Dice  & mean IoU & $F_{\beta}^{w}$ & $\quad S_{\alpha}$ & $E_{\phi}^{\max }$ & MAE   \\ \hline 

    Circle & \underline{$0.913$} & \underline{$0.849$} &\underline{$0.925$} &\underline{$0.929$} &\underline{$0.980$} &\underline{0.008} \\
    Triangle & $0.888$ & $0.811$ &$0.903$ &$0.911$ &$0.963$ &$0.010$ \\
    Square & $0.876$ & $0.789$ &$0.894$ &$0.900$ &$0.958$ &$0.012$ \\
    Pentagon & $0.857$ & $0.765$ &$0.874$ &$0.889$ &$0.947$ &$0.013$ \\
    \textbf{Ours} & \textbf{0.940} & \textbf{0.890} & \textbf{0.949} & \textbf{0.950} & \textbf{0.988} & \textbf{0.008} \\
    
    \hline
\end{tabular}    
}
    \vspace{0.3cm}\caption{Comparative results of using different geometric shapes.}
    \label{table:ablation_study_2}
\end{table}

\begin{figure}[t]
\begin{center}
  \includegraphics[width=1\linewidth]{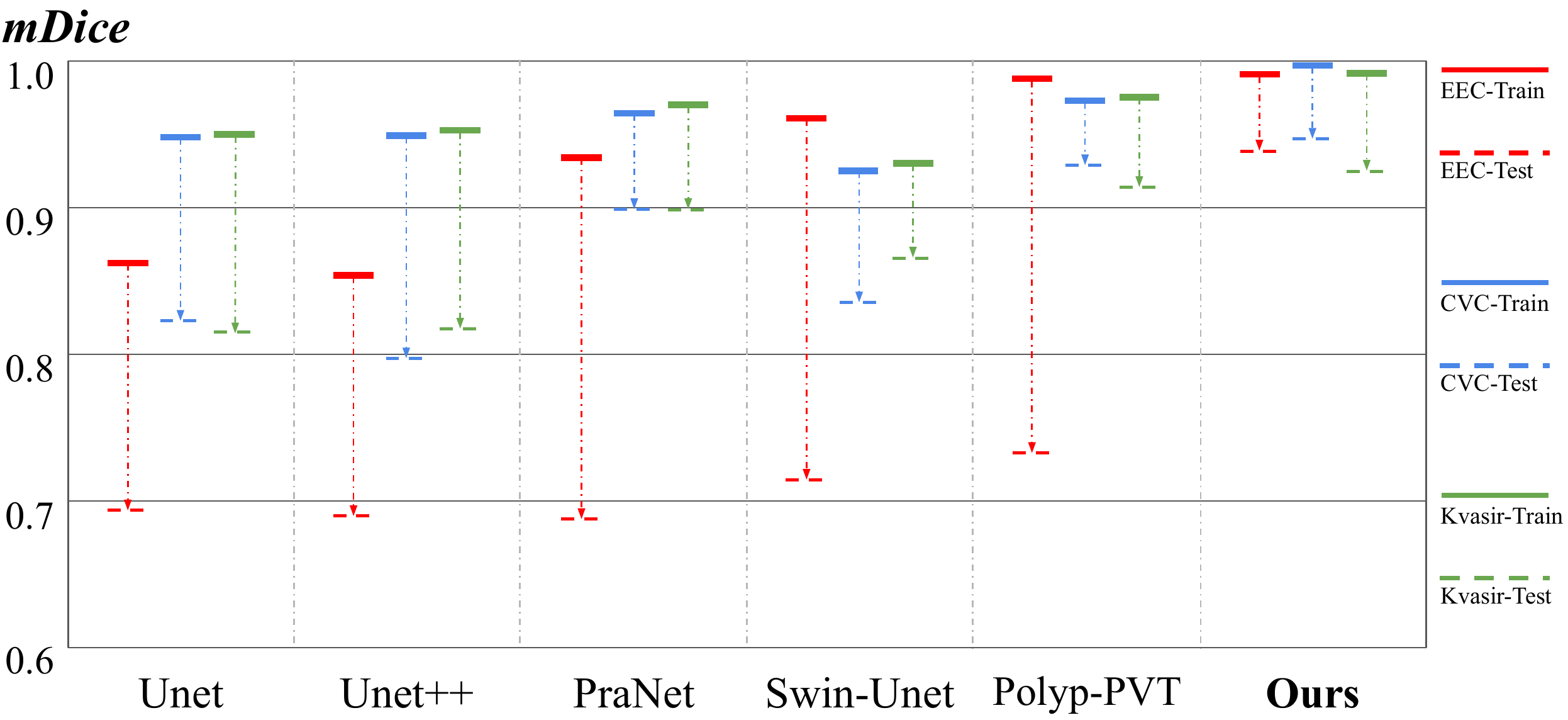}
\end{center}
  \caption{Training accuracy versus testing accuracy of various methods over three datasets.}
\label{fig:train_acc}
\end{figure}

\begin{figure}[h]
\begin{center}
  \includegraphics[width=1\linewidth]{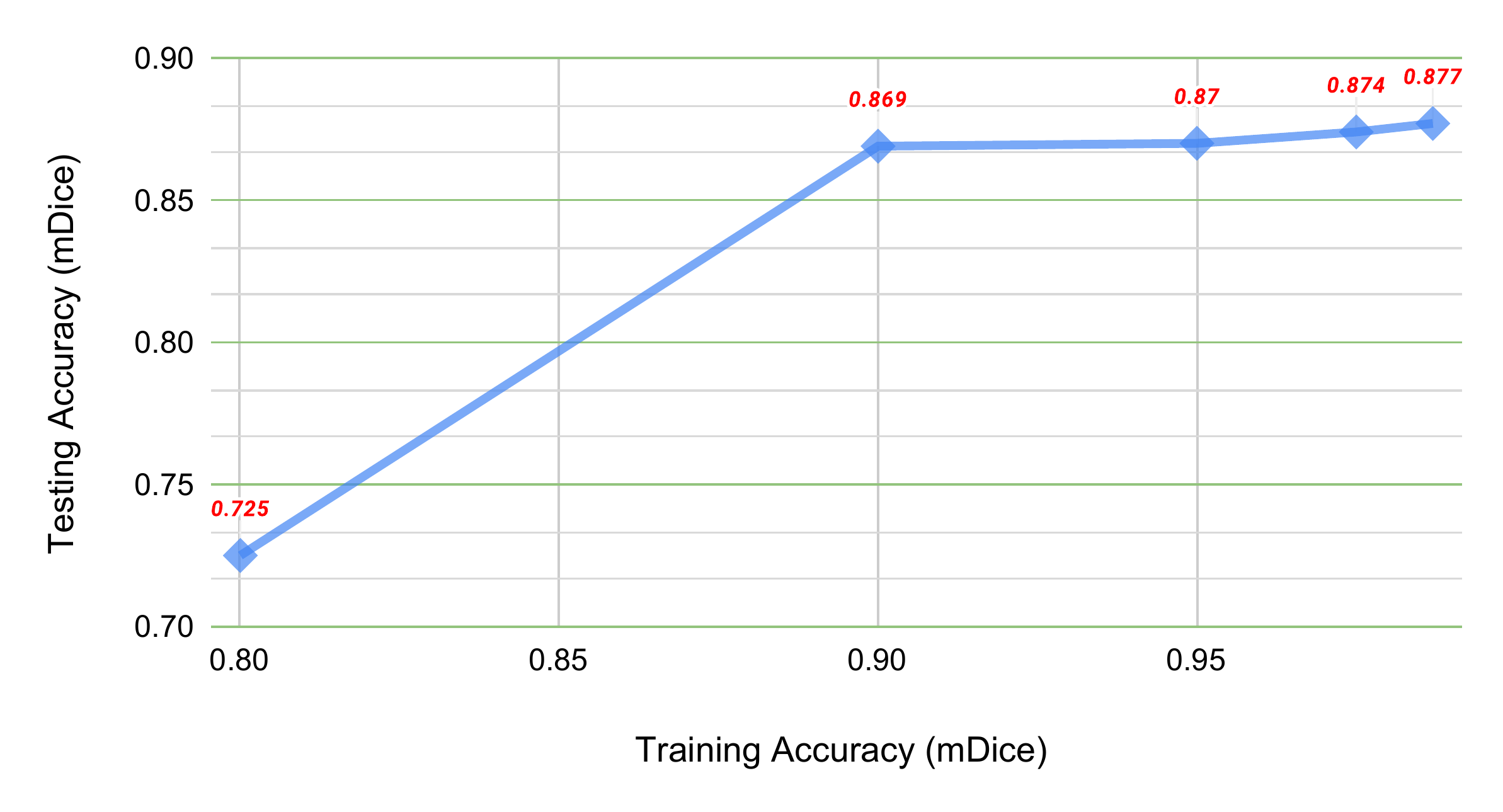}
\end{center}
  \caption{Test accuracy versus training accuracy of eUNet over EEC-2022.}
\label{fig:eUnet_train_test_acc}
\end{figure}

\subsection{Comparison with SAM.} SAM is the recent "segment anything model", proposed most recently by Meta~\cite{kirillov2023segment}. It is trained on a large data-set of 11 million licensed images and over 1 billion masks, and achieves SOTAs by a large margin in multiple segmentation tasks. Here, we use it to segment EEC lesions where both the maximum outer rectangle of the ROI regions and the sampling points are fed into SAM as prompts. The first row of Fig.~\ref{sam_teaser} shows the segmentation results. For the purpose of comparison, the second row of Fig.~\ref{sam_teaser} shows the corresponding results obtained by YOHO. Clearly, YOHO outperforms SAM by a large margin in terms of the achieved Dice scores.

\subsection{Ablation Studies}
\label{sec:ablation}

The first ablation study focuses on the effectiveness of the proposed edge block and the corresponding loss. It can be seen from the results shown in Table~\ref{table:ablation_study_3} that estimating an accurate edge map and enabling the edge consistency loss indeed improves the results.

The second ablation study is designed to investigate the effect of using different geometric shapes, including circles,  triangles, squares, and pentagons, in the data augmentation process. It can be seen from Table~\ref{table:ablation_study_2} that the resulted performance ranks from high to low following the order of circles, triangles, squares, and pentagons. Therefore, the best combination is to use circles and triangles, as we have adopted in our YOHO framework.

The last experiment, which is more significant, is related to the “over-fitting” issue. To this goal, we record the training accuracy and the test accuracy of each network and draw them in vertical lines with bars on top and at bottom, as shown in Fig.~\ref{fig:train_acc}. Here, the top bar and bottom bar of each line represent the training accuracy and the test accuracy, respectively. Several observations can be made as follows. (1) Over two polyp datasets, each chosen benchmark network maintains a high training accuracy while its test accuracy drops only by an acceptable amount. In particular, the drop is small in three more sophisticated networks, i.e., PraNet, Swin-UNet, and Polyp-PVT. This indicates that the ``over-fitting" problem is minimal. (2) Over the EEC dataset, however, PraNet, Swin-UNet, and Poly-PVT are suffering seriously from the “over-fitting” problem. Although the training accuracy remains at a very high level, the test accuracy drops by about 25 percentage points. (3) The test accuracy of our approach drops from the training accuracy by a smaller amount in all three datasets. Finally, the test accuracy versus the training accuracy of our eUNet over EEC-2022 is plotted in Fig.\ref{fig:eUnet_train_test_acc} to further prove that our method indeed leverages the “over-fitting” characteristic.

\section{Conclusion}
\label{sec:conclusion}
To our best knowledge, the ``one-image-one-network" solution for the lesion segmentation in endoscopy images proposed in this paper is original. The resulted YOHO framework brings up two big advantages: (1) no image datasets need be collected from multiple patients so as to provide a complete protection of patients' privacy and (2) the so-called ``over-fitting" problem, perhaps the most critical generalization-related problem, is now solved elegantly. Technically, our work also makes two major contributions. First, the proposed geometry-based rendering of a single input image to generate the training set seems to be the first attempt in the field. Second, the EEC dataset we create is likely to play an important role in promoting future researches. Experimental results over two polyp datasets and one EEC dataset provide a strong evidence that YOHO represents the state-of-the-art (SOTA). Our future works include fine-tuning of the YOHO framework, further speeding up the training, and applying the similar idea to other types of medical images such as CT and MRI images. Code and dataset will be released to facilitate future researches.

{
\small
\bibliographystyle{ieeetr}
\bibliography{yoho}
}

\end{document}